\documentclass[sigconf]{acmart}
\usepackage{graphicx} 
\usepackage{xcolor}
\usepackage{listings}
\usepackage{balance}
\usepackage{enumitem}
\usepackage{array}

\usepackage{flushend}

\usepackage{listings}

\definecolor{codebg}{RGB}{245,245,244}

\usepackage{placeins}

\usepackage{subcaption} 
\usepackage{tabularx}
\usepackage{array}
\definecolor{codebg}{RGB}{245,245,245}
\usepackage{hyperref}
\usepackage{soul}
\usepackage[normalem]{ulem}

\newenvironment{compact_enum}{
  \begin{itemize}[leftmargin=*, itemsep=0pt, parsep=2pt, topsep=2pt]
}{
  \end{itemize}
}

\lstdefinestyle{custom}{
  basicstyle=\ttfamily\small,
  backgroundcolor=\color{white},
  keywordstyle=\color{blue},
  commentstyle=\color{gray},
  showstringspaces=false,
  breaklines=true,
  moredelim=**[is][\colorbox{yellow}]{@}{@}
}

\newcolumntype{A}{>{\hsize=3\hsize\centering\arraybackslash}X}
\newcolumntype{B}{>{\hsize=2\hsize\centering\arraybackslash}X}


\title{A Unified Structured Query Understanding Framework for Industrial Semantic Search}

\author{Ping Liu}
\affiliation{
 \institution{LinkedIn Corporation}
 \city{Mountain View}
 \state{CA}
 \country{USA}
}
\email{piliu@linkedin.com}

\author{Qianqi Shen}
\affiliation{
 \institution{LinkedIn Corporation}
 \city{Mountain View}
 \state{CA}
 \country{USA}}
\email{qishen@linkedin.com}

\author{Jianqiang Shen}
\affiliation{
 \institution{LinkedIn Corporation}
 \city{Mountain View}
 \state{CA}
 \country{USA}}
\email{jershen@linkedin.com}

\author{Chunnan Yao}
\affiliation{
 \institution{LinkedIn Corporation}
 \city{Mountain View}
 \state{CA}
 \country{USA}}
\email{chyao@linkedin.com}

\author{Kevin Kao}
\affiliation{
 \institution{LinkedIn Corporation}
 \city{Mountain View}
 \state{CA}
 \country{USA}}
\email{kkao@linkedin.com}

\author{Rajat Arora}
\affiliation{
 \institution{LinkedIn Corporation}
 \city{Mountain View}
 \state{CA}
 \country{USA}}
\email{rajarora@linkedin.com}

\author{Dan Xu}
\affiliation{
 \institution{LinkedIn Corporation}
 \city{Mountain View}
 \state{CA}
 \country{USA}}
\email{dnxu@linkedin.com}

\author{Baofen Zheng}
\affiliation{
 \institution{LinkedIn Corporation}
 \city{Mountain View}
 \state{CA}
 \country{USA}}
\email{bzheng@linkedin.com}

\author{Yunxiang Ren}
\affiliation{
 \institution{LinkedIn Corporation}
 \city{Mountain View}
 \state{CA}
 \country{USA}}
\email{yunren@linkedin.com}

\author{Benjamin Hoan Le}
\affiliation{
 \institution{LinkedIn Corporation}
 \city{Mountain View}
 \state{CA}
 \country{USA}
}
\email{ble@linkedin.com}

\author{Ali Hooshmand}
\affiliation{
 \institution{LinkedIn Corporation}
 \city{Mountain View}
 \state{CA}
 \country{USA}
}
\email{ahooshmand@linkedin.com}

\author{Igor Lapchuk}
\affiliation{
 \institution{LinkedIn Corporation}
 \city{Mountain View}
 \state{CA}
 \country{USA}
}
\email{ilapchuk@linkedin.com}

\author{Juan Bottaro}
\affiliation{
 \institution{LinkedIn Corporation}
 \city{Mountain View}
 \state{CA}
 \country{USA}
}
\email{jbottaro@linkedin.com}

\author{Raghavan Muthuregunathan}
\affiliation{
 \institution{LinkedIn Corporation}
 \city{Mountain View}
 \state{CA}
 \country{USA}
}
\email{rmuthuregunathan@linkedin.com}

\author{Caleb Johnson}
\affiliation{
 \institution{LinkedIn Corporation}
 \city{Mountain View}
 \state{CA}
 \country{USA}}
\email{cajohnson@linkedin.com}

\author{Liangjie Hong}
\affiliation{
 \institution{LinkedIn Corporation}
 \city{Mountain View}
 \state{CA}
 \country{USA}}
\email{liahong@linkedin.com}

\author{Jingwei Wu}
\affiliation{
 \institution{LinkedIn Corporation}
 \city{Mountain View}
 \state{CA}
 \country{USA}}
\email{jingwu@linkedin.com}

\author{Wenjing Zhang}
\affiliation{
 \institution{LinkedIn Corporation}
 \city{Mountain View}
 \state{CA}
 \country{USA}}
\email{wzhang@linkedin.com}

\copyrightyear{2026}
\acmYear{2026}
\setcopyright{cc}
\setcctype{by}
\acmConference[KDD '26]{Proceedings of the 32nd ACM SIGKDD Conference on Knowledge Discovery and Data Mining V.2}{August 09--13, 2026}{Jeju Island, Republic of Korea}
\acmBooktitle{Proceedings of the 32nd ACM SIGKDD Conference on Knowledge Discovery and Data Mining V.2 (KDD '26), August 09--13, 2026, Jeju Island, Republic of Korea}
\acmDOI{10.1145/3770855.3818312}
\acmISBN{979-8-4007-2259-2/2026/08}

\begin{document}

\begin{abstract}

Query understanding in large-scale industrial search systems is typically implemented as a cascade of disparate, task-specific components. While individually optimizable, this fragmented architecture incurs high maintenance overhead and results in inconsistent behaviors, particularly for long-tail queries. In this work, we propose and deploy a unified structured query understanding system that consolidates these heterogeneous functions into a single Small Language Model (SLM) that performs schema-constrained generation. To address the data bottlenecks inherent in unified modeling, we introduce \textit{Query Illuminator}, a dual-purpose framework serving as: (i) a teacher model for high-quality auto-annotation and distillation, and (ii) a surrogate judge for scalable evaluation where human labels are scarce. We validate this approach through extensive offline and online tests within LinkedIn's Job Search system. Furthermore, we demonstrate the framework's horizontal extensibility through a cross-domain case study on People Search. The results show improved user engagement and reduced operational costs, achieved while satisfying strict low-latency serving constraints on limited GPU resources.

\end{abstract}


\maketitle

\section{Introduction}
Industrial semantic search systems must interpret variable-length, ambiguous user queries and convert them into reliable signals that downstream retrieval and ranking components can act on. In many vertical search products, the \emph{query understanding} layer is implemented as a pipeline of separately trained models and rules (e.g., entity tagging and query rewriting). Such fragmentation increases maintenance cost, makes behavior inconsistent across query types, and amplifies errors in the long tail.

We study this problem within LinkedIn’s large-scale job search system, which indexes  $>10^7$ documents and handles traffic $>10^4$ queries per second (QPS). Compared to general web search, job search introduces domain-specific constraints such as (i) a two-sided marketplace with inventory liquidity, (ii) member-context dependence (e.g., location, skills, seniority), and (iii) strict trust requirements. These factors are common across industrial semantic search verticals, but are especially pronounced in job search.

Traffic in \emph{Semantic Job Search} (SJS) is distinct from the legacy keyword-oriented \emph{Classical Job Search} (CJS) by a heavy tail of high-entropy natural-language inputs. Although both systems share a median query length of 2 words, SJS frequently encounters verbose requests that expose the brittleness of fragmented pipelines (Table~\ref{tab:sjs_sample_queries}). For example, queries like \textit{``easy apply remote data entry clerk''} intertwine functional UI constraints with semantic roles; \textit{``salary $>X$ or remote in US''} requires joint reasoning over numerical and geo constraints. These complex dependencies often fail to resolve consistently in cascaded architectures, where isolated components risk over-extracting entities or missing latent intents.

Motivated by these constraints, we propose \emph{unified structured query understanding}: instead of maintaining separate predictors and taggers, we use a single Small Language Model (SLM) \cite{schick2021s} to generate all query understanding artifacts in a schema-constrained output. Formally, given a query $q$ and context $c$, the model produces
\[
y=\{r,t,w,s,\ldots\}=f(q,c),
\]
where $r$ is the router output (query-type / operation indicator), $t$ are structured tags, $w$ is an optional rewrite, and $s$ is an optional trust/safety decision. Crucially, routing ($r$) is generated as part of the same structured output, in lieu of an external intent classifier.

Another challenge is that unified modeling exacerbates data and evaluation bottlenecks. Query understanding labels are expensive: correctness depends not only on semantic plausibility but also on product policy (e.g., ambiguity handling, avoidance of hallucinated filters, and strict trust rules). Furthermore, many failure modes are subtle and cannot be captured by standard automatic metrics.
To address this, we introduce \emph{Query Illuminator}, a high-capacity offline framework that serves both as a teacher model for large-scale auto-annotation and knowledge distillation, and as an LLM-based judge enabling task-agnostic offline evaluation. In large-scale online A/B tests, the unified system delivers consistent relevance gains while satisfying strict serving-time constraints (Section~\ref{sec.infra}).

\begin{table}[tb]
\caption{Sanitized examples of Semantic Job Search queries sampled and the corresponding patterns.}
\vspace{-6pt}
\centering
\small
\setlength{\tabcolsep}{4pt}
\begin{tabularx}{\columnwidth}{@{}X>{\raggedright\arraybackslash}p{0.34\columnwidth}@{}}
\toprule
\textbf{Sanitized SJS query samples} & \textbf{Illustrated pattern} \\
\midrule
AI engineer & short role/title \\
summer 20XX internship & temporal constraint \\
entry level geotechnical engineer & long-tail multi-token role \\
remote in \texttt{[STATE]} & geo + remote intent \\
easy apply remote data entry clerk & UI-specific intent + remote + role \\
find oncology clinical development roles with salary $>$ \texttt{\$[AMOUNT]} in \texttt{[CITY]} or remote in US posted in past \texttt{[N]} weeks & long-form multi-constraint request \\
\bottomrule
\end{tabularx}
\vspace{-2pt}
\label{tab:sjs_sample_queries}
\end{table}

This paper presents a production query understanding framework that consolidates routing, tagging, rewriting, facet suggestion, and trust handling into a single schema-constrained SLM designed for low-latency industrial search. Our contributions are:
\begin{compact_enum}
\item \textbf{Unified structured query understanding.} We formulate query understanding as schema-constrained generation, replacing fragmented multi-stage pipelines with a single SLM and reducing error propagation.
\item \textbf{Policy-grounded supervision and evaluation.} We introduce a large teacher model for scalable auto-annotation and distillation, and a policy-grounded LLM judge with a lightweight robustness validation protocol.
\item \textbf{Scalable serving for structured generation.} We describe a production serving stack with guided decoding, schema validation, and batching, and report load-tested latency and request-level failure behavior under realistic QPS.
\item \textbf{Large-scale deployment results.} Online A/B tests show statistically significant improvements in engagement metrics (Table~\ref{tab:AB_combined}), validating the approach in production.
\item \textbf{Cross-domain extensibility.} We demonstrate portability to Semantic People Search (Section~\ref{sec.sps}) with minimal adaptation to a different schema and executors.
\end{compact_enum}

\begin{figure*}[tb]
    \centering
    \includegraphics[width=0.996\textwidth]{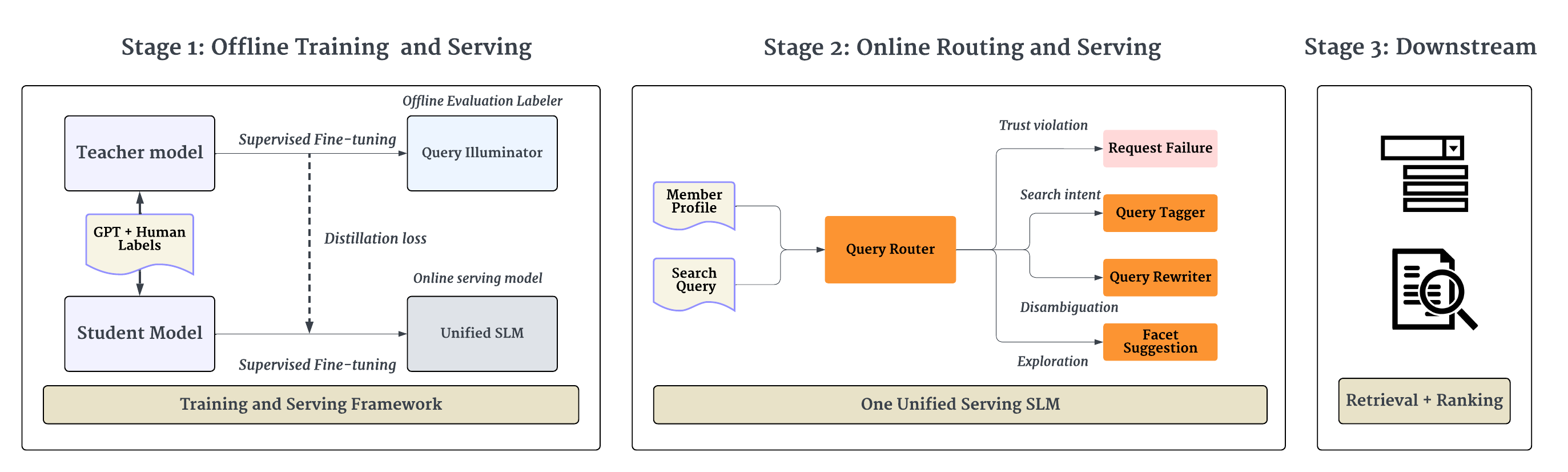}
    
    \caption{The workflow of our query understanding framework. The components in the center are powered by a single unified SLM. The outputs generated by the query understanding module are consumed by downstream applications.}
  
    \label{fig.workflow}
    \Description[The pipeline of framework.]{The pipeline of framework.}
\vspace{6pt}
\end{figure*}

\section{Related Work}
\label{sec.related_work}

\textit{\textbf{Query Understanding Pipelines and Selective Transformations}} Prior to the rise of generative AI, query understanding was typically organized as a set of modular stages, including query parsing or segmentation, semantic tagging, query expansion, and query rewriting, with the goal of enriching raw user queries to improve retrieval effectiveness \cite{jones2006generating, li2022query, robertson1995okapi, selvaretnam2012natural}. For instance, \citet{jones2006generating} leverage search logs to mine query modifications, Microsoft \cite{wang2015query} aligns query terms with ontology concepts to enable query conceptualization, and Taobao \cite{li2022query} investigates query rewriting through retrieval-and-rank augmentation. Collectively, these efforts help explain why production query understanding systems in industry often develop into collections of independently optimized components. A related line of work focuses on \emph{selective transformation}: while rewriting or expansion can benefit certain query intents, it may negatively impact others, which has motivated research on query performance prediction and selective reformulation \cite{cronen2002predicting, carmel2010estimating, he2007qpp}. Our approach adopts this selective perspective but replaces multiple standalone triggers with a unified structured prediction framework, in which the Query Router jointly predicts a discrete routing decision $r$ and other structured outputs, including tags, optional rewrites, facet triggers, and trust-related decisions.


\textit{\textbf{LLM-based Query Understanding in Production. }} As large language models have grown more capable, prior work has increasingly examined generative methods for query understanding and reformulation. QUILL \cite{srinivasan2022quill} enhances query interpretation using retrieval-augmented context, while BEQUE \cite{peng2024large} introduces a staged query rewriting framework for Taobao search. Related studies include query-understanding-driven multi-task learning for ranking in Amazon search \cite{luo2024exploring} and reinforcement learning approaches to improving reformulation quality \cite{agrawal2023enhancing}. More recent results indicate that even relatively compact SLMs can achieve strong performance on query understanding tasks \cite{dhole2024genqrensemble}. We frame the \emph{query understanding layer} as a unified production interface that jointly predicts routing decisions and structured artifacts within a single schema-constrained output, with production evidence of robustness and request-level failure handling.



\textit{\textbf{Structured Outputs and Serving-time Reliability. }} Deploying generative models in industrial search requires predictable, structured outputs and stable serving behavior under strict latency budgets. Recent systems therefore increasingly leverage constrained decoding, schema validation, and optimized serving stacks, including guided grammar decoding \cite{dong2024xgrammar} and high-throughput LLM serving \cite{kwon2023vllm}. Our system adopts schema-constrained function calling (tool invocation with typed names and JSON arguments) and guided decoding to reduce malformed outputs, and we characterize latency behavior (mean/median/P95/P99) under load. We use tool interfaces as a \emph{structured output format} in a single turn, rather than an open-ended multi-step tool-use setting \cite{yao2023react, schick2023toolformer}.

\textit{\textbf{Teacher Distillation and LLM-based Evaluation. }} Large teacher models \cite{shi2025enhancing,yuan2025hardness,wang2025put} are commonly used to produce synthetic supervision and distill their capabilities into smaller, deployable models \cite{hinton2015distilling, alpaca}. In the meantime, LLM-based judges have been adopted as surrogate evaluators in settings where human labels are scarce \cite{zheng2023mtbench, liu2023geval, dubois2023alpacaeval, wang2023ultrafeedback}, though they are known to introduce biases and therefore require explicit validation. In our system, Query Illuminator serves both functions: acting as a teacher for auto-annotation and distillation, and as a judge for surrogate evaluation. Section~\ref{sec.experiment} describes a lightweight judge validation protocol that calibrates and bounds surrogate judge signals prior to their use in major decision-making.


Our work is orthogonal to embedding-based retrieval and LLM ranking models \cite{juan2025scalingretrieval, li2025mixlm}, focusing instead on the upstream query understanding interface that generates structured signals consumed by retrieval and ranking components.

\begin{table}[t]
\centering
\footnotesize
\caption{\textbf{Key differences from our prior work~\cite{liu2025powering}.}}
\setlength{\tabcolsep}{3pt}
\begin{tabular}{p{2cm} p{2cm} p{3.2cm}}
\toprule
\textbf{Aspect} & \textbf{Prior Work} & \textbf{This Work} \\
\midrule

Training 
& SFT 
& \textbf{Distillation + auto-labeling + active learning} \\

Evaluation 
& Offline + A/B 
& \textbf{LLM judge (Illuminator) + human validation} \\

Generalization 
& Job Search 
& \textbf{+ Semantic People Search (+57\% / +75\%)} \\

Serving 
& Basic deployment 
& \textbf{Grammar decoding + batching + caching} \\

\bottomrule
\end{tabular}
\label{tab:short_diff}
\vspace{-3pt}
\end{table}

Compared to our previous work on query understanding~\cite{liu2025powering}, this paper extends the unified query understanding framework along several key dimensions: training, evaluation, and generalization. First, we introduce a knowledge distillation pipeline using a high-capacity Query Illuminator, enabling a compact 1.5B model to achieve strong performance under strict latency constraints. Second, we develop a dual-purpose policy-grounded framework where the same system serves as both a teacher for supervision and a structured LLM-based judge for scalable evaluation. Third, we demonstrate cross-domain generalization by extending the framework from Semantic Job Search to Semantic People Search, achieving substantial gains in graded relevance. In addition, we provide significantly expanded serving and latency analysis, including batching, prefix caching, and schema-constrained decoding optimizations. Finally, we augment evaluation with multilingual benchmarks, human studies, and large-scale A/B testing, offering a more comprehensive validation of the system in production. Together, these contributions transform the prior work into a more scalable, extensible, production-complete, and closed-loop system for industrial semantic search.

\section{Unified Framework Architecture}
\label{sec.framework}
We propose a unified framework for structured query understanding that reconciles the expressive semantic requirements of industrial search with the strict latency constraints of online serving. The architecture decouples these competing objectives into two asynchronous environments: a low-latency \textit{Online Inference Engine} powered by a specialized SLM, and a high-capacity offline \textit{Query Illuminator} that leverages LLMs for supervision synthesis and policy-grounded evaluation. 



\subsection{Online Inference: Structured Generation}
Figure~\ref{fig.workflow} illustrates the end-to-end system topology.  The core of the online engine is a monolithic SLM optimized for single-pass, schema-constrained generation. Formally, given a user query $q$ and a context vector $c$ (comprising member profile, location, and session history), the model estimates the joint probability of a structured artifact $y$:
\[
y = \{r, t, w, s, \ldots\} \sim P_\theta(y \mid q, c)
\]
where the output schema is strictly enforced during decoding. The router token $r$ acts as a latent control variable, gating the generation of subsequent fields such as entity tags ($t$), query rewrites ($w$), or safety classifications ($s$).

Unlike cascading pipelines or multi-turn agentic systems which incur compounding latency and risk error propagation between disjoint modules, our unified formulation enforces a holistic decision process. By generating routing decisions and artifacts in a single forward pass, the model can capture inter-dependencies (e.g., conditioning a rewrite on the detected intent) while guaranteeing deterministic latency bounds essential for downstream retrieval and ranking.

While our primary deployment targets job search, the schema-centric formulation is inherently domain-agnostic; extending the system to new verticals requires only redefining the output schema, as demonstrated in the Semantic People Search case study (Section~\ref{sec.sps}), which operates over $>10^9$ documents and $>10^4$ QPS.

\subsection{Query Illuminator: Offline Reasoning}
To address supervision scarcity and the challenge of evaluating open-ended structured generation, we introduce \textit{Query Illuminator}, an offline framework that operationalizes product policies into computational signals using high-capacity LLMs. The Illuminator serves two tightly coupled roles: supervision synthesis and policy-aligned evaluation.

\textbf{Teacher for Supervision Synthesis.}
First, the Illuminator functions as a teacher model $T$ for training the student SLM $S$. Given an input tuple $(q, c)$, $T$ generates high-fidelity, schema-compliant targets $y$, including routing decisions, tags, and rewrites. This teacher–student distillation process transfers reasoning capabilities from a high-capacity model into a compact, deployable SLM (e.g., Qwen2.5-1.5B~\cite{qwen2.5}), effectively decoupling reasoning complexity from online inference latency.

\textbf{Judge for Policy-Aligned Evaluation.}
Second, the Illuminator acts as a surrogate evaluator. In industrial contexts, ``correctness'' is defined not merely by semantic plausibility but by strict adherence to product policies such as conservative handling of ambiguity and rigid anti-hallucination constraints. We encode these requirements via two versioned prompt templates:
\begin{compact_enum}
    \item \textit{Labeling prompts}, which produce schema-constrained targets for supervised fine-tuning and distillation;
    \item \textit{Evaluation prompts}, which implement a critic function $E(y_{\text{pred}}, y_{\text{ref}})$ that compares candidate outputs against references, assigns rubric-based scores (e.g., extraction precision, query hygiene), and emits a structured error taxonomy for automated analysis.
\end{compact_enum}

\textbf{The Policy-Grounded Loop.}
We version the policy specification, golden sets, and prompts together, and calibrate prompt instructions on human-labeled examples until rubric-aligned outputs are achieved. Section~\ref{sec.judge} describes how we validate the robustness of this surrogate-judge protocol (e.g., JSON parse stability across versions) and how we bound judge-based claims during development.
Finally, live traffic is continuously sampled and rescored by the Illuminator in a daily offline pipeline, providing near real-time monitoring of distribution shifts and a principled mechanism for continual learning.

\subsection{Query Router}
The query router corresponds to the router output $r$ in our structured output $y=\{r,t,w,s,\ldots\}$. In general, the router \cite{mu-etal-2024-query,kang2022study} takes one of a small set of discrete values that gate downstream executors (e.g., tag extraction, profile-aware rewrite, facet suggestion, or trust blocking). Because $r$ is generated as part of the same schema-constrained output, routing is trained jointly with tagging/rewriting/facet/trust decisions and does not require a separate intent-classification stage.

\begin{figure}[tb]
\begin{center}
\begin{minipage}{\columnwidth}
\centering 
\begin{lstlisting}
You are a helpful assistant designed to prepare job search requests for the user. Please only process information that is explicitly mentioned in the search query.

You are provided a list of tools that you can invoke to better assist with the user.

interface QueryEngine {
    job_in_network_tool?: {
    /**
    * Processes the job search query if jobs within the user's network are requested. DO NOT use this tool if the query does not explicitly mention information related to user's network connections.
    */
    includeOrExclude: boolean; // True if the query specifies jobs within the user's network, false if the query specifies jobs out of user's network.
    };
}
\end{lstlisting}
\vspace{-6pt}
\captionof{figure}{An example of prompt snippet for processing job search requests based on network inclusion criteria.}
\label{fig.prompt}
\end{minipage}
\end{center}
\vspace{0pt}
\end{figure}

In our deployment we use four coarse routing types:

\begin{compact_enum}
    \item \textbf{Criteria queries:} extract structured constraints (e.g., title, location, company, work modality) and optionally trigger facet suggestion.
    \item \textbf{Self-referential queries:} trigger query rewriting using member context (e.g., ``near my location'', ``jobs I'm qualified for'') before downstream retrieval.
    \item \textbf{Unsupported or unclear intent:} default to a high-recall, keyword-only lexical retrieval path as a conservative fallback to mitigate the risk of over-structuring.
    \item \textbf{Trust-violating queries:} block or deflect requests that violate platform trust/safety policies.
\end{compact_enum}

\subsection{Query Tagger}
Query tagging \cite{tang2025llm4tag} converts free-text job queries into structured constraints that downstream retrieval and ranking can consume (e.g., title, company, location, seniority, and request-level intents such as ``remote'' or ``Easy Apply'') \cite{manshadi2009semantic, zuo2023deeptagger}. Our approach leverages prompt-based instruction and enforces structured output formats (e.g., JSON, TypeScript), allowing the model to flexibly and accurately extract job-related entities. Unlike traditional encoder-based or NER systems that depend on static vocabularies, our LLM architecture can infer implicit relationships and reason over contextual cues, thereby improving robustness to long-tail or novel entity types. The model effectively handles both core attributes, such as company, title, and location, and extended request-level attributes like ``Easy Apply'', application count limits, and remote or hybrid work preferences.

These extracted attributes directly enhance multiple downstream components of the recommendation pipeline. Certain facets, such as location and ``Easy Apply'', are used as filters during the candidate selection stage to narrow the retrieval space efficiently. Meanwhile, text-based attributes, such as job title, serve as semantic features that enrich both retrieval and ranking models, improving relevance and personalization. 
We evaluate the downstream impact of our framework via an offline analysis within an Embedding-Based Retrieval (EBR) system~\cite{juan2025scalingretrieval}. Integrating our structured tagging results into the EBR pipeline without retraining demonstrates substantial performance improvements in search relevance, as detailed in Section~\ref{exp:ebr}.


\subsection{Query Rewriter}
Query rewriter \cite{feng2025complicated,shao2025great,wang2025personalized} refines a user’s search query to enhance the relevance of retrieved jobs. It disambiguates the implicit intent by integrating information from the user’s profile and historical job activities, clarifying both the semantic meaning of the query and the underlying intent.

For self-referential queries, the rewriter injects member context (e.g., location or profile attributes) to produce an explicit query string for retrieval. When such self-reference is detected, the LLM processes both the original query and the member profile as inputs. For example, the query ``Find software engineer jobs near my location'' may be rewritten as ``Find software engineer jobs near Bay Area, CA'' if the member’s profile indicates that location. The rewritten query is then forwarded to the candidate selection and ranking modules, resulting in more accurate and contextually relevant matches.

\subsection{Query Facet Suggestion}
\begin{figure}[t!]
    \centering
    \includegraphics[width=0.32\textwidth,height=5.5cm]{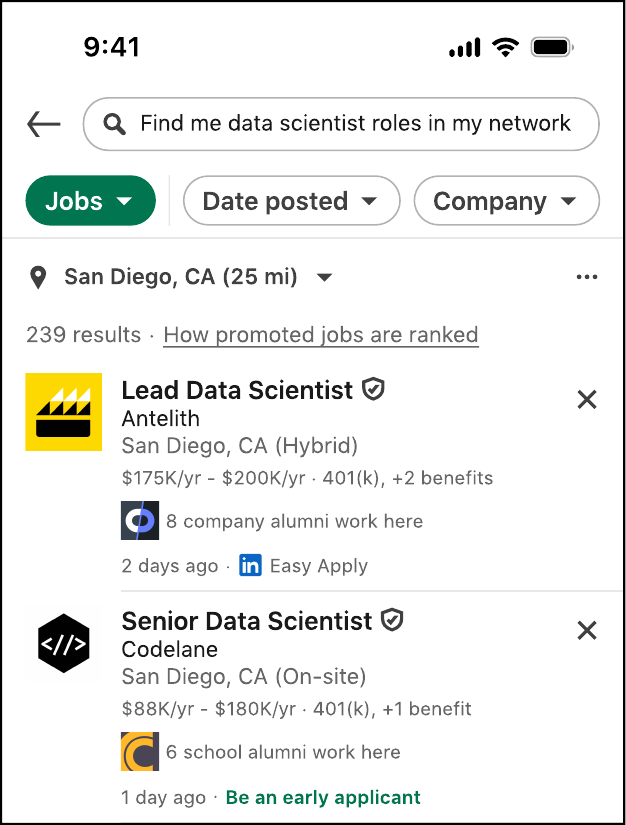}
    \caption{A sample UI generated by the query engine, as displayed in the interface. The facet is represented by the round button located beneath the input search field. }
    \label{fig.UI}
    \Description[The pipeline of framework.]{The pipeline of framework.}
\end{figure}

Facet suggestion recommends query-relevant filters (e.g., location, company, seniority) that help users refine and explore results (Figure~\ref{fig.UI}). We construct candidate facets by normalizing historical query entities into facet types and retrieving a small query-relevant subset via a lightweight embedding index. The Unified SLM then selects and scores facets under a schema-constrained interface. Facet suggestion is gated by the router output $r$ to avoid distracting suggestions for intents where facets are unlikely to help.

To maintain user trust, our framework is designed around the principle that semantic query understanding should expose, rather than obscure, the underlying search affordances. The model generates explicit, typed search constraints (e.g., Company, Location), which are immediately surfaced in the user interface (Figure~\ref{fig.UI}). These inferred facets remain fully editable, allowing semantic parsing to operate in tandem with traditional explicit filtering rather than replacing it. The framework exposes an interpretable and collaborative search state, enabling transparent user control at production scale.

By consolidating all query understanding capabilities into a unified LLM framework, our system deprecates multiple legacy NER pipelines, minimizes maintenance costs, and improves the factuality and scalability of the entire query understanding stack.

\section{Modeling Methodology}
\label{sec.model}
This section details the dataset construction procedure and the two-stage training paradigm, consisting of Supervised Fine-Tuning (SFT) and Knowledge Distillation (KD), which forms the foundation of our unified framework.

 \subsection{Data Construction \& Active Learning}

We construct an instruction-tuning dataset consisting of $\sim$14K production queries, each annotated with schema-constrained structured outputs. To ensure label fidelity and policy compliance, we adopt a human-in-the-loop bootstrapping pipeline: candidate annotations are first generated by the \textit{Query Illuminator}, then reviewed by expert annotators to enforce policy alignment. In addition, to bridge the gap between historical keyword centric paradigm and semantic queries we aim to power, we transform raw keyword queries using an Alpaca‑style \cite{alpaca} instruction‑generation pipeline that rewrites each query into a natural‑language form while triggering a randomized set of tool invocations and their associated responses, yielding a richer instruction‑tuning dataset.
To improve robustness under long-tail distribution shift, we further implement a \textit{continuous active learning loop}. Each day, we sample live traffic and rescore it offline using the Illuminator. We prioritize samples that are informative for model improvement, including cases with high teacher--student disagreement and queries exhibiting novel semantic patterns.

No personally identifiable information (PII) was used in any stage of model training. All datasets were anonymized and aggregated to ensure compliance with privacy and data protection standards~\cite{hall2013differential,gao2019exploring}.

\subsection{Training Strategy}
We train a compact production model via multi-task supervised fine-tuning (SFT) on schema-constrained outputs, followed by distillation from a larger Query Illuminator. The unified model supports two output modes: (i) schema-constrained structured generation for routing/tags/facets/safety and (ii) optional natural-language rewriting when triggered by the router.


\subsubsection{Supervised fine-tuning}
We adopt 1.5B language model as the production backbone and a larger Query Illuminator (8B language model) offline.  We formulate the query understanding task as conditional generation, optimizing the model parameters $\theta$ to minimize the next-token prediction cross-entropy loss:


\begin{equation}
\mathcal{L}_{\text{SFT}} = -\sum_{i=1}^N \log P_\theta(y_i \mid x_i),
\end{equation}

where $x_i$ is the input prompt (query + context + schema), $y_i$ is the target structured output, $P_\theta(y_i \mid x_i)$ is the probability of $y_i$ given $x_i$ under model parameters $\theta$, and $N$ is the total number of training examples.

\subsubsection{Distillation fine-tuning}

To bridge the reasoning gap between the 1.5B Student and the 8B Teacher, we employ Knowledge Distillation \cite{hinton2015distilling}. The Teacher generates probability distributions (soft targets) that encode uncertainty and semantic relationships between classes—information lost in discrete ``hard'' labels, thereby achieving stronger generalization and stability under limited labeled data.
Formally, the distillation objective combines the supervised fine-tuning loss with a Kullback–Leibler (KL) divergence term between the teacher and student output distributions. The total loss is expressed as:

\begin{equation}
\mathcal{L}_{\text{total}} = \mathcal{L}_{\text{SFT}} + \lambda \, \mathcal{L}_{\text{KD}},
\end{equation}

where

\begin{equation}
\mathcal{L}_{\text{KD}} = \frac{1}{N} \sum_{i=1}^{N} D_{\text{KL}}\!\left( 
P_{\theta_T}(y \mid x_i) \,\|\, P_{\theta_S}(y \mid x_i)
\right),
\end{equation}

with $P_{\theta_T}$ and $P_{\theta_S}$ denoting the teacher and student output distributions, respectively, and $\lambda$ controlling the relative weight between imitation and supervised learning. In practice, reasoning-based distillation often requires a smaller $\lambda$ while non-reasoning tasks tend to benefit from a larger $\lambda$.

\subsubsection{Multi-task Fine-tuning}

We adopt a multi-task instruction tuning paradigm~\cite{sanh2021multitask}, optimizing a single SLM across heterogeneous tasks using shared parameters. This unification allows the model to learn a shared output manifold for all objectives.
Joint optimization induces \textit{positive transfer}. High-level intent recognition (e.g., routing) and fine-grained entity extraction (e.g., tagging) rely on intrinsically coupled semantic features. Training concurrently forces the model to internalize a coherent latent representation of query semantics, preventing the fragmented learning typical of isolated models.
Furthermore, multi-tasking acts as a \textit{regularizer}, as detailed in Appendix A. Single-task models often overfit to dominant lexical patterns (e.g., memorizing keywords). The requirement to satisfy diverse, competing objectives constrains the hypothesis space, promoting generalizable inference over brittle heuristics. Section~\ref{sec.experiment} validates this: multi-task models outperform specialists on tail queries (Table~\ref{table:single_multi_tasks}), demonstrating that task heterogeneity prevents over-specialization to head distributions.


\begin{table}[tb]
\caption{Stretched benchmark results of latency}
\vspace{-7pt}
\centering
\textit{Query Tagging Throughput \& Latency Benchmark}
\vspace{4pt}
\begin{tabular}{@{\hskip 5pt}l@{\hskip 10pt}c@{\hskip 5pt}c@{\hskip 5pt}c@{\hskip 5pt}c@{\hskip 5pt}}
\toprule
\textbf{QPS} & \textbf{1} & \textbf{5} & \textbf{10} & \textbf{15} \\
\midrule
Median E2E Latency (ms) & 243.47 & 261.25 & 261.24 & 275.81 \\
P90 E2E Latency (ms)    & 366.98 & 388.64 & 398.89 & 407.65 \\
P95 E2E Latency (ms)    & 431.83 & 467.12 & 433.28 & 441.57 \\
P99 E2E Latency (ms)    & 553.58 & 588.59 & 605.89 & 625.93 \\
\bottomrule
\end{tabular}
\vspace{3pt}

\textit{Facet Suggestion Throughput \& Latency Benchmark}
\begin{tabular}{@{\hskip 5pt}l@{\hskip 10pt}c@{\hskip 5pt}c@{\hskip 5pt}c@{\hskip 5pt}c@{\hskip 5pt}}
\toprule
\textbf{QPS} & \textbf{1} & \textbf{5} & \textbf{10} & \textbf{15} \\
\midrule
Median E2E Latency (ms) & 109.66 & 104.35 & 102.08 & 106.30 \\
P90 E2E Latency (ms)    & 115.15 & 110.49 & 107.07 & 116.55 \\
P95 E2E Latency (ms)    & 117.23 & 112.69 & 108.99 & 124.49 \\
P99 E2E Latency (ms)    & 121.95 & 123.92 & 393.35 & 566.46 \\
\bottomrule 
\end{tabular}
\par
\vspace{-6pt}
\label{tab.benchmarks_aligned}
\end{table}

\section{Infrastructure and Serving}
\label{sec.infra}
\subsection{Inference Serving System}

Our serving stack is designed to meet two competing goals: (i) low end‑to‑end latency at production traffic, and (ii) high throughput under bursty workloads, while preserving flexibility for prompt/model iteration. At a high level, the online server comprises vLLM~\cite{kwon2023vllm} wrapped by a hosted LLM service, which is called by upstream application backend services gated by a cache layer.


We employ single-turn decoder-only LLMs exposed through an OpenAI-compatible completions interface. Since number of tools are finite and fixed we use guided grammar with xgrammar \cite{dong2024xgrammar} to speed up inference. To further improve production robustness, we implement tolerant JSON parsing and recovery logic. In cases where the generated output is partially incomplete or contains minor structural deviations due to token limits or generation errors, the application layer attempts schema-aware correction and recovery rather than failing the request outright.

Facet suggestion performs point-wise scoring over 20--30 candidate facets per request. Client-side batching is implemented in the application tier before calling the hosted model which avoids repeated HTTP/TCP overheads, adapter resolution, and tokenizer warm-up that would otherwise occur for each sub-request under server-side batching. Client‑side batching reduces p95 latency by over 70\% compared to server-side batching at the same QPS.

Prefix caching further improves inference efficiency. In our setting, the system prompt is shared across all requests, the member profile is shared within a session, and only the search query varies per request. By placing shared components at the beginning of the prompt, prefix caching avoids redundant pre-filling. When the member profile dominates the prompt, a dummy warm-up request can be issued at the start of each batch to cache it, allowing subsequent requests to skip profile pre-filling. This optimization trades additional memory and computation for the warm-up request against reduced per-request latency.

\subsection{Benchmark}
Table~\ref{tab.benchmarks_aligned} reports a stretched throughput/latency benchmark for two tasks in our unified serving stack: Query Tagging and Facet Suggestion. We replay 1{,}000 sampled production queries through vLLM~\cite{kwon2023vllm} on a single NVIDIA A100 MIG-4g.40G GPU and sweep offered load from 1 to 15 QPS. 
Query Tagging is a single schema-constrained generation (input $\sim$2000 tokens, output 20--100) with median latency less than 300 ms. Facet Suggestion ranks candidate facets via point-wise scoring implemented as a client-side batch of short decoding requests, achieves 110 ms as median latency. Comparing two tables, query tagging scales more gracefully due to bounded decoding space, while facet suggestion degrades earlier due to combinatorial expansion. Across both tasks, latency remains within strict production requirements under constrained GPU resources.

\section{Experimental Results}
\label{sec.experiment}
We evaluate the unified query understanding layer along four axes: (i) offline quality of structured outputs (including multilingual extraction and an MT fine-tuning ablation), (ii) analysis and discuss how we leverage query illuminator in the production as LLM judge, (iii) online impact in A/B tests on job-search surfaces, and (iv) portability and downstream consumption via a cross-domain people-search case study and a retrieval integration example.

\subsection{Query Length Analysis}
To contextualize the operating regime for semantic job search, we compute query-length statistics (whitespace-delimited words) on U.S. traffic from 2025. Table~\ref{tab:query_length_dist} compares Semantic Job Search (SJS) and Classical Job Search (CJS), each split by Free vs.\ Premium members. While the median query length is two words across segments, SJS exhibits a heavier tail: Premium SJS reaches 5 words at the 90th percentile and 18 words at the 99th percentile. This heavier tail amplifies long-tail entities and compositional constraints, motivating robust structured generation with explicit validation and fallbacks. For comparison, when SJS ramped to 1\% of global members, the average query length was 2.74 words per query.

\begin{table}[tb]
\caption{Distribution of query length}
\vspace{-6pt}
\centering
\small
\setlength{\tabcolsep}{5pt}
\begin{tabular}{lrrrr}
\toprule
 & \multicolumn{2}{c}{CJS} & \multicolumn{2}{c}{SJS} \\
\cmidrule(lr){2-3}\cmidrule(lr){4-5}
Statistic (words/query) & Free & Premium & Free & Premium \\
\midrule
Average           & 2.3 & 2.5 & 2.7 & 3.3 \\
50th percentile   & 2.0 & 2.0 & 2.0 & 2.0 \\
90th percentile   & 3.0 & 3.0 & 4.0 & 5.0 \\
99th percentile   & 8.0 & 11.0 & 11.0 & 18.0 \\
\bottomrule
\end{tabular}
\label{tab:query_length_dist}
\end{table}

\subsection{Offline Evaluation}

We evaluate Query Tagger on an internal multilingual dataset covering English, German, Spanish, and French. Table~\ref{table.offline} reports precision/recall across model variants, including a distilled 1.5B student trained from the Query Illuminator teacher.
The results illustrate two practical lessons for industrial query understanding. First, scaling alone is not strictly monotonic: the 7B model increases English recall but can reduce precision compared to the 1.5B baseline, reflecting the precision-recall trade-off of generative extraction. Second, distillation is an effective path to production: the 1.5B distilled student recovers much of the larger teacher's quality across languages while retaining a smaller serving footprint.

\begin{table}[tb]
\caption{Precision and recall for multilingual query tagging.}
\vspace{-6pt}
\centering
\small
\setlength{\tabcolsep}{4pt}
\resizebox{\columnwidth}{!}{%
\begin{tabular}{lcccccccc}
\toprule
Model & \multicolumn{2}{c}{English} & \multicolumn{2}{c}{German} &
\multicolumn{2}{c}{Spanish} & \multicolumn{2}{c}{French} \\
\cmidrule(lr){2-3}\cmidrule(lr){4-5}\cmidrule(lr){6-7}\cmidrule(lr){8-9}
& Prec. & Rec. & Prec. & Rec. & Prec. & Rec. & Prec. & Rec. \\
\midrule
qwen2.5-1.5b-sft       & 0.823 & 0.843 & 0.910 & 0.900 & 0.902 & 0.944 & 0.933 & 0.969 \\
qwen2.5-7b-sft         & 0.815 & 0.892 & 0.918 & 0.949 & 0.929 & 0.962 & 0.946 & 0.978 \\
qwen3-8b-sft           & 0.867 & 0.857 & 0.937 & 0.953 & 0.925 & 0.957 & 0.957 & 0.978 \\
qwen2.5-1.5b-distill   & 0.857 & 0.878 & 0.909 & 0.956 & 0.929 & 0.957 & 0.957 & 0.969 \\
\bottomrule
\end{tabular}}
\label{table.offline}
\end{table}

We additionally conduct a human evaluation on company extraction, comparing the previous production system, a GPT-based baseline, and the Query Illuminator teacher (Table~\ref{tab:metric_comparison}).

\begin{table}[tb]
\caption{Human evaluation on company tool}
\vspace{-7pt}
\centering
\begin{tabular}{lccc}
\hline
\textbf{Metric} & \textbf{Baseline} & \textbf{GPT} & \textbf{Illuminator} \\
\hline
Simple Accuracy           & 84.00\%  & 90.50\%  & 91.50\%  \\
Recall (TPR)              & 72.62\%  & 96.55\%  & 95.12\%  \\
False Alarm Prob. (FPR)   & 7.76\%   & 14.16\%  & 11.02\%  \\
Precision                 & 87.14\%  & 84.00\%  & 85.71\%  \\
F-measure                 & 79.22\%  & 89.84\%  & 90.17\%  \\
\hline
\end{tabular}
\vspace{-6pt}
\label{tab:metric_comparison}
\end{table}

\textbf{Multi-task vs.\ single-tool fine-tuning ablation.}
To isolate the effect of multi-task instruction tuning, we benchmark the unified model against a single-task baseline trained exclusively on the Location extraction tool. Table~\ref{table:single_multi_tasks} shows that joint optimization yields significant gains: multi-task fine-tuning improves Location Precision/Recall from 0.899/0.871 to 0.928/0.919, and lifts overall F1 from 0.885 to 0.924. We attribute this to \textit{positive transfer}: the shared representation learning across heterogeneous tasks acts as a regularizer, preventing the model from overfitting to the narrow heuristics of a single tool. 

As a preliminary diagnostic of general-purpose capability, we evaluate both variants on the GLUE benchmark~\cite{wang2018glue}. The results suggest a specialization trade-off resembling an ``alignment tax''~\cite{ouyang2022training}: the multi-task model yields modest gains on semantic similarity and grammatical acceptability tasks, potentially due to paraphrase-sensitive representations induced by the rewriting objective, but shows regressions on natural language inference tasks, which may indicate objective interference.
Overall, these suggest that multi-task tuning improves the model’s utility as a structured semantic parser for tool invocation, while potentially reducing broad inference generalization. This trade-off is acceptable for industrial query understanding, where schema adherence and reliable downstream control take precedence over open-ended deductive reasoning.



\begin{table}[tb]
\caption{Accuracy of location tools under single-tool vs.\ multi-tool fine-tuning (same backbone and training setup).}
\vspace{-2pt}
\centering
\small
\setlength{\tabcolsep}{6pt}
\begin{tabular}{lcccc}
\toprule
\textbf{Setting} & \textbf{Prec.} & \textbf{Rec.} & \textbf{F1} & \textbf{\shortstack{Parameter\\Accuracy}} \\
\midrule
Location only & 0.899 & 0.871 & 0.885 & 0.908 \\
Location with other tools & 0.928 & 0.919 & 0.924 & 0.928 \\
\bottomrule
\end{tabular}
\vspace{-6pt}
\label{table:single_multi_tasks}
\end{table}

\subsection{Surrogate LLM Judge Validation}\label{sec.judge}
High-quality evaluation is a bottleneck for structured query understanding: many failure modes are subtle (e.g., hallucinated structured filters, incorrect facet typing, or overly aggressive disambiguation), and manual review does not scale across iterative model versions. We therefore use the Query Illuminator as a surrogate judge, but ground it in explicit product policy to reduce drift.

We validate this protocol on Semantic People Search, which maps people-search queries into a fixed schema of supported facets (e.g., person name, company, school, location, network distance) under strict anti-hallucination and ambiguity handling. We curate a golden set of 4{,}903 production queries with human-labeled structured outputs. Given (i) the raw query, (ii) the human reference, and (iii) the system output, the judge returns two rubric-based sub-scores --- \emph{facet extraction} and \emph{query hygiene} --- graded as Great/Fair/Poor (mapped to 2/1/0) and summed into an overall graded-relevance (GR) score on a 0--4 scale. It also emits a structured mistake taxonomy (e.g., ambiguity errors, over/under-extraction, entity-type confusion, and canonicalization/spellcheck issues) that supports debugging and regression tracking. 

Since our framework relies heavily on Query Illuminator as a surrogate evaluator for rapid offline iteration and large-scale assessment, establishing its reliability is critical. We conducted an internal validation study on approximately 300K samples. The Illuminator–human agreement achieved Cohen’s $\kappa=0.77$, compared to a human-expert agreement ceiling of 0.83. This strong agreement indicates that the evaluator provides a stable and high-fidelity approximation of expert judgment, making it suitable as an evaluation signal for large-scale development and model iteration.

As a robustness check for automation, we require the judge to emit strictly valid JSON; across prompt/model iterations on the same golden set, JSON parse success ranges from 99.96\% to 100\%. Aggregated GR scores provide a stable development signal (average 2.95--3.25 on the 0--4 scale across iterations), while the taxonomy enables targeted data augmentation. We treat judge-based scores as an auxiliary development signal rather than ground truth: prompts and rubrics are versioned, and we periodically audit disagreement cases as product policy evolves.  

\subsection{Online A/B Evaluation}
We conducted large-scale randomized A/B experiments on two primary production surfaces: \textit{Semantic Job Search} (SJS) and \textit{Job Alert}. Traffic was gradually ramped from a dark launch to a 1\% exposure and ultimately rolled into a 50/50 control-versus-treatment split over a 14-day observation window, to detect distributional failures early. Significance was determined using two-sided $t$-tests with a threshold of $p < 0.05$. The control group operated on the legacy heterogeneous pipeline, consisting of: (1) a decoder-only Query Engine~\cite{liu2025powering} responsible for intent decoding and blocking, (2) an encoder-only BERT-based tagger fine-tuned for a restricted set of entity types (primarily Company and Location).
Table~\ref{tab:AB_combined} summarizes the significant metric changes with p-value less than 0.05. 

In SJS, the unified system reduces request failures by 17.13\% while improving engagement proxies such as apply-to-viewport ratio (+0.14\%), search sessions (+0.74\%), and successful apply-or-save actions (+0.49\%). In Job Alert, we observe consistent positive movement on alert engagement. In production, when query understanding layer incorrectly classifies a valid query as unsupported (a false positive block) or fails to produce an executable plan (due to tagging), the request becomes a failure that prevents retrieval. The unified system reduces request failures by 17.13\%, which is mainly driven by improvements in routing and tagging. This reliability gain coincided with improvements in top‑level engagement and outcome metrics. We attribute these shifts to more reliable structured routing and tagging that better supports downstream retrieval and ranking. Additionally, we track the schema compliance rate as a primary health metric to monitor model stability.

\begin{table}[tb]
\caption{Relative metrics of the online A/B test.}
\vspace{-2pt}
\centering
\small
\setlength{\tabcolsep}{6pt}
\begin{tabular}{l l r}
\toprule
Category & Metric & Relative change \\
\midrule
Job Search & Apply-to-viewport ratio   & $+0.14\%$ \\
Job Search & Request failures          & $-17.13\%$ \\
Job Search & Search sessions           & $+0.74\%$ \\
Job Search & Successful apply or save  & $+0.49\%$ \\
\midrule
Job Alert  & Total job applications    & $+0.23\%$ \\
Job Alert  & Job search sessions       & $+0.07\%$ \\
Job Alert  & Searches without results  & $-0.65\%$ \\
Job Alert  & Impressions at ranks 1--5 & $+0.10\%$ \\
\bottomrule
\end{tabular}
\vspace{-4pt}
\label{tab:AB_combined}
\end{table}

\subsection{Cross-domain: Semantic People Search}\label{sec.sps}
To assess the portability beyond job search, we apply the same unified structured query understanding interface to Semantic People Search (SPS). SPS uses a different schema (person-centric and organization-centric entities) and different executors, but reuses the same serving stack (schema-constrained generation, validation, and routing via $r$).

Early in SPS ramp, real query logs are sparse; we therefore augment fine-tuning with templated synthetic queries instantiated from a knowledge graph. We evaluate SPS using the policy-grounded surrogate judge protocol from Section~\ref{sec.judge}.
We retain a lightweight first-pass router to separate classical name lookups from semantic SPS queries. In a U.S.\ ramp, 70.1\% of traffic is routed to the semantic SPS path and 26.2\% to the classical path (Table~\ref{tab:sps_routing_mix}); 3.7\% falls back to a safe path and $<0.1\%$ is blocked for trust/safety.

\begin{table}[tb]
\caption{People-search routing categories and distribution.}
\vspace{-4pt}
\centering
\small
\setlength{\tabcolsep}{4pt}
\begin{tabularx}{\columnwidth}{@{}l r r X@{}}
\toprule
Router result & Count & Share & Notes \\
\midrule
SPS\_SUPPORTED        & 1{,}091{,}876 & 70.1\% & Semantic people-search pipeline (schema constrained QU + retrieval). \\
CPS\_SUPPORTED        &   408{,}486 & 26.2\% & Classical people search (name-only). \\
ROUTER\_UNSUPPORTED   &    57{,}879 &  3.7\% & Out-of-scope intent; routed to a safe fallback. \\
ROUTER\_NEGATIVE      &       279 & $<0.1$\% & Blocked for trust/safety. \\
\bottomrule
\end{tabularx}
\vspace{-0pt}
\label{tab:sps_routing_mix}
\end{table}

\paragraph{\textbf{End-to-end quality impact via offline graded relevance.}}
We estimate end-to-end impact using a production replay on a held-out golden query set: control sends raw queries to retrieval/ranking, while treatment consumes the QD-decoded structured query. A high-capacity teacher judge scores the top-10 results per query with graded relevance (0--4). Table~\ref{tab:sps_gr_qd} reports large relative lifts: +57\% on navigational queries and +75\% on exploratory queries.

\begin{table}[tb]
\caption{Relative lift in offline graded relevance (GR@10) when SPS retrieval consumes QD-decoded structured queries.}
\vspace{-4pt}
\centering
\small
\setlength{\tabcolsep}{4pt}
\begin{tabularx}{\columnwidth}{@{}X r@{}}
\toprule
Bucket & Rel. change \\
\midrule
Navigational queries & +57\% \\
Exploratory queries & +75\% \\
\bottomrule
\end{tabularx}
\vspace{-0pt}
\label{tab:sps_gr_qd}

\end{table}

\subsection{Downstream Consumption in Retrieval}
\label{exp:ebr}

As an example of downstream consumption, we inject structured outputs~\cite{zhang2024automated} from the unified query engine into an existing embedding based retrieval (EBR) system~\cite{juan2025scalingretrieval}. We do not propose a new retrieval algorithm; instead, we show that explicit structured constraints can be incorporated into prompt/feature construction. Many search queries are domain-specific yet ambiguous in general language (e.g., company names that overlap with common words). For instance, the query ``Anthropologie'' may refer to the U.S.\ retailer or to the French word for ``anthropology''. When the Query Tagger identifies an unambiguous company constraint, we augment the EBR prompt with a dedicated \texttt{Job Company} field (Figure~\ref{fig.sjs_prompt} in  Appendix~\ref{sec.appendix}). In an offline evaluation, this simple injection (without EBR retraining) improves the top-50 retrieval metrics by +1.8\% precision, +2.2\% recall, and +1.7\% NDCG shown in Table~\ref{table.ebr_perf}.

\begin{table}[tb]
\caption{EBR top-50 retrieval metrics with and without adding a structured company tag to the prompt (Figure~\ref{fig.sjs_prompt}).}
\vspace{-4 pt}
\centering
\small
\setlength{\tabcolsep}{6pt}

\begin{tabular}{lccc}
\toprule
\textbf{Setting} & \textbf{Prec@50} & \textbf{Rec@50} & \textbf{NDCG@50} \\
\midrule
Baseline                   & 0.577 & 0.699 & 0.687 \\
+ Company tag in prompt & \textbf{0.595} & \textbf{0.721} & \textbf{0.704} \\
\bottomrule
\end{tabular}
\vspace{-0pt}
\label{table.ebr_perf}

\end{table}

\section{Discussion and Limitations}\label{sec.discussion}

\textbf{Context Selection and Inference Efficiency.} In industrial semantic search, effective query understanding relies on conditioning the model not only on the raw input but also on auxiliary context (e.g., user profiles, job descriptions, and policy documents). However, unconditional retrieval of candidate documents introduces a trade-off: excessive context increases inference latency and risk of `lost-in-the-middle'' hallucination, while insufficient context leads to constraint omission. To mitigate this, we restrict input to a curated set of high-density features and utilize the router output $r$ to gate computationally expensive operations. More principled, learning-based evidence selection is a promising direction.

\textbf{Query Ambiguity and Implicit Intent.} Short-text queries (1--2 words) constitute a significant portion of production traffic. These inputs frequently exhibit high entropy, lacking explicit constraints regarding location, seniority, or work modality. While our model leverages user-specific context to infer latent intent, there is an upper bound on non-interactive resolution. In cases of irreducible ambiguity, the optimal strategy shifts from inference to interactive disambiguation (e.g., conversational slot-filling).  We consider interactive UX design as outside the scope of this paper.

\textbf{Schema Rigidity and Domain Generalization.} The design of our framework deliberately leverages the stability of structured output schemas to guarantee deterministic downstream execution. While exploring environments with rapid taxonomy drift remains an active area of research, our current framework delivers immediate, scalable utility for core enterprise domains. The empirical success of this approach is supported by the SPS case study (Section~\ref{sec.sps}), which showcases successful infrastructure reuse across disparate schemas. Ultimately, this work offers a highly repeatable blueprint for unified query understanding.

\textbf{Limitations of Surrogate Evaluation.} We employ a teacher LLM (Query Illuminator) as a scalable surrogate evaluator to approximate human judgment. As detailed in Section~\ref{sec.judge}, we mitigate the risk of metric divergence through policy-anchored prompting and robustness checks (e.g., JSON syntax validation). Nevertheless, surrogate metrics remain susceptible to drift if the evaluation prompt diverges from evolving product policies. We address this by version-controlling prompts and periodically auditing disagreement rates, with plans to rigorously quantify the correlation between judge and human annotations in future work.

\section{Conclusion}\label{sec.conclusion}

In this work, we presented the architectural design, operationalization, and production deployment of a unified structured query understanding framework for large-scale industrial semantic search. The system consolidates routing, tagging, rewriting, facet suggestion, and trust/safety handling into a single schema-constrained Small Language Model. To scale development under label scarcity and evolving product policies, we introduced the Query Illuminator used offline as both a high-capacity teacher for synthetic supervision/distillation and a surrogate judge for evaluation. 
In large-scale online experiments in job search, the system reduced request failures by 17.13\% with positive engagement shifts, and load tests show the serving stack meets strict latency budgets. A cross-domain case study on semantic people search demonstrates that our framework can be reused with a different schema and executors. Future work includes deeper empirical validation of the surrogate judge's alignment with human experts, advanced contextual evidence selection, and the development of interactive, glass-box UX interfaces for disambiguating open-ended or underspecified queries.

\newpage

\section*{Acknowledgement}

We thank the many talented engineers and collaborators at LinkedIn whose contributions made this work possible. We also thank the anonymous reviewers for their thoughtful and constructive feedback, which helped improve the quality and clarity of this paper.

\bibliographystyle{ACM-Reference-Format}
\bibliography{main}

\section*{Appendix}
\label{sec.appendix}

\subsection*{A. Multi-task Fine-tuning as Capability Retention}
We include a brief interpretation of multi-task fine-tuning that motivates its use as a practical regularizer in our structured tool setting.

\paragraph{\textbf{Proposition A.1 (Multi-task objective as a relaxation of capability-retention constraints).}}
Let $L_0(\theta)$ denote the loss for a primary task and $L_k(\theta)$ for $k=1,\ldots,K$ denote losses for auxiliary tasks to be retained. Consider the constrained problem:
\begin{equation}
\min_{\theta}\; L_0(\theta)\quad \text{s.t.}\quad L_k(\theta)\le \varepsilon_k\;\; \forall k=1,\ldots,K.
\end{equation}

The associated Lagrangian is

\begin{equation}
\mathcal{L}(\theta,\lambda)=L_0(\theta)+\sum_{k=1}^{K}\lambda_k\big(L_k(\theta)-\varepsilon_k\big),\quad \lambda_k\ge 0.
\end{equation}

For any fixed $\lambda\ge 0$, minimizing $\mathcal{L}(\theta,\lambda)$ over $\theta$ is equivalent (up to an additive constant independent of $\theta$) to minimizing the weighted multi-task objective:
\begin{equation}
L_0(\theta)+\sum_{k=1}^{K}\lambda_k L_k(\theta).
\end{equation}
\textit{Proof sketch.} The term $-\sum_k \lambda_k\varepsilon_k$ does not depend on $\theta$ and thus is constant with respect to $\theta$. Under standard convexity assumptions and Slater’s condition, strong duality implies there exists $\lambda^\star$ such that a primal optimum also minimizes the Lagrangian at $\lambda^\star$. While deep nets are non-convex, this provides a useful interpretation: multi-task fine-tuning can be viewed as a relaxation of ``retain auxiliary capabilities'' constraints.

\paragraph{\textbf{Corollary A.2 (Local quadratic view: multi-task as a structured regularizer).}}
Let $\theta_0$ be the pretrained initialization. If each auxiliary loss admits a local second-order approximation near $\theta_0$, then
\begin{equation}
L_k(\theta)\approx L_k(\theta_0)+\tfrac{1}{2}(\theta-\theta_0)^\top H_k(\theta-\theta_0),
\end{equation}
then the multi-task objective is approximately
\begin{equation}
L_0(\theta)+\tfrac{1}{2}(\theta-\theta_0)^\top\Big(\sum_{k=1}^{K}\lambda_k H_k\Big)(\theta-\theta_0)+\text{const},
\end{equation}
which is equivalent to optimizing the primary loss with an additional quadratic penalty that discourages moving far from $\theta_0$ in directions that increase auxiliary losses. This formalizes the intuition that joint tool training can act as a practical regularizer supporting capability retention.

\subsection*{B. Semantic People Search Analysis}

SPS queries are short on average but exhibit a non-trivial tail: while the median length is 2 words, the 99th percentile falls in the 8+ word bucket. Table~\ref{tab:sps_examples} shows representative sanitized queries and the corresponding structured patterns. 

\begin{table}[tb]
\caption{Representative sanitized People Search queries and the corresponding structured patterns.}
\vspace{-6pt}
\centering
\small
\setlength{\tabcolsep}{4pt}
\begin{tabularx}{\columnwidth}{@{}X X@{}}
\toprule
Sanitized query & Structured pattern (illustrative) \\
\midrule
\texttt{[PERSON\_NAME]} & \texttt{person\_name=[PERSON\_NAME]} \\
\texttt{[PERSON\_NAME] at [COMPANY]} & \texttt{person\_name=[PERSON\_NAME], current\_company=[COMPANY]} \\
\texttt{[COMPANY] CEO} & \texttt{current\_company=[COMPANY], keywords="CEO"} \\
\texttt{former [COMPANY] employees} & \texttt{past\_company=[COMPANY]} \\
\texttt{[SCHOOL] alumni data science} & \texttt{school=[SCHOOL], keywords="data science"} \\
\texttt{customer success leader} & \texttt{keywords="customer success leader"} \\
\bottomrule
\end{tabularx}
\label{tab:sps_examples}
\end{table}

\begin{table}[tb]
\caption{Illustrative SPS structured output schema.}
\vspace{-6pt}
\centering
\small
\setlength{\tabcolsep}{4pt}
\begin{tabularx}{\columnwidth}{@{}lX@{}}
\toprule
Field & Description \\
\midrule
person\_name      & Person name if explicitly specified. \\
current\_company  & Constraint on current employer / organization. \\
past\_company     & Constraint on prior employer / organization. \\
school            & Education constraint (school / university). \\
keywords          & Additional free-text constraints (skills, titles, locations). \\
rewrite           & Optional rewritten query used for retrieval. \\
\bottomrule
\end{tabularx}
\label{tab:sps_schema}
\end{table}

Table~\ref{tab:sps_schema} summarizes the structured output schema used by the SPS component. Table \ref{tab:sps_qd_serving} reports the serving capacity of the Query Decoder (QD) under different model sizes and hardware configurations, measured in queries per second (QPS) at a p95 latency constraint below 500 ms. The baseline QD-1.7B model achieves approximately 12 QPS on a full A100 GPU, establishing a feasible production reference point. Simply upsizing the model to QD-4B on the same hardware is not viable due to latency and throughput constraints. Moving the 4B model to a more powerful H100 GPU restores feasibility and increases throughput to around 20 QPS. Finally, applying prompt compression on H100 further improves efficiency, boosting capacity to roughly 30 QPS. These results highlight the combined importance of hardware selection and input optimization in enabling larger models to meet strict serving latency requirements.

\begin{table}[H]
\caption{Query Decoder (QD) serving capacity in SPS.}
\vspace{-6pt}
\centering
\small
\setlength{\tabcolsep}{4pt}
\begin{tabular}{lcc}
\toprule
Configuration & GPU & QPS at p95$<$500ms \\
\midrule
QD-1.7B (baseline) & A100 (full) & $\sim$12 \\
QD-4B (upsize) & A100 (full) & not feasible \\
QD-4B (upsize) & H100 (full) & $\sim$20 \\
QD-4B + prompt compression & H100 (compressed) & $\sim$30 \\
\bottomrule
\end{tabular}
\label{tab:sps_qd_serving}
\end{table}

\balance

\subsection*{C. Additional Detail of Data Collection}

For routing and tagging tasks, model outputs were formatted as structured tool-calling responses. These outputs specify which downstream task or module the query should be routed to, and which tags best capture its semantic and functional attributes. However, using job-search interaction logs as-is creates a chicken-and-egg problem: historical queries were written in a keyword-centric paradigm and therefore do not reflect the semantic queries we want to enable in job search. To address this mismatch, we adopt two strategies. (1) We extract raw keyword queries and user-selected facets from the classical search interface, then prompt GPT to rewrite them into natural-language queries that implicitly express the corresponding facet operations. (2) We create synthetic instruction following data similar to \cite{alpaca} in which GPT is prompted to transform a given query into one that explicitly invokes a random selection of tool calls and additionally outputs the corresponding tagger response. To increase coverage and robustness, we also randomly inject company names and geographic locations during query generation. For query rewriting and facet suggestion tasks, contextual relevance plays a crucial role. To simulate realistic personalization, we synthetically generated member profiles and corresponding job-seeking activities for each query. Both the query and the member context were then provided as input to the model—either to produce refined query reformulations or to suggest complementary facets (e.g., industry, company, or location) aligned with the user’s inferred intent.

\begin{figure}[tb]
  \centering
  \begin{subfigure}[t]{0.49\linewidth}
    \centering
\begin{lstlisting}
Read the following job seeker query and focus on the user's requirements including job title, occupation, skill, specialty, location, company, and compensation range.

Think about how this information might be relevant for potential job for this job seeker.

Query: <QUERY>

\end{lstlisting}
    \caption{Query Prompt}
    \label{fig:sjs_prompt:query}
  \end{subfigure}\hfill
  \begin{subfigure}[t]{0.49\linewidth}
    \centering
\begin{lstlisting}
Read the following job seeker query and focus on the user's requirements including job title, occupation, skill, specialty, location, company, and compensation range.

Think about how this information might be relevant for potential job for this job seeker.
Query: <QUERY>
Job Company: <COMPANYNAME>
\end{lstlisting}
    \caption{Query prompt adding structured company name}
    \label{fig:sjs_prompt:query_company}
  \end{subfigure}
  \caption{Enhancement of EBR module prompts by adding tagged company names}
  \label{fig.sjs_prompt}
\end{figure}

\begin{table}[tb]
\caption{Evaluation metrics of facet suggestion for different fine-tuning strategies.}
\vspace{-6pt}
\centering
\small
\setlength{\tabcolsep}{4pt}
\begin{tabularx}{\columnwidth}{Xccc}
\toprule
Fine tune strategy & Eval Loss & Precision & F1-score \\
\midrule
Teacher Model (SFT) & 0.3751 & 0.79 & 0.81 \\
Student Model (SFT) & 0.3896 & 0.69 & 0.77 \\
Distill & 0.3643 & 0.70 & 0.78 \\
SFT + Distill & 0.3428 & 0.75 & 0.81 \\
\bottomrule
\end{tabularx}
\label{tab:finetune_eval}
\end{table}

\subsection*{D. Additional Offline Evaluation}

Table~\ref{tab:finetune_eval} evaluates different fine-tuning strategies for facet suggestion across three key metrics: evaluation loss, precision, and F1-score. While the standalone Student Model (SFT) underperforms compared to the Teacher Model across all metrics, applying distillation yields notable improvements. Specifically, the pure "Distill" strategy successfully lowers the evaluation loss to 0.3643 and slightly edges out the student model's F1-score. However, the hybrid "SFT + Distill" approach delivers the best overall performance, achieving the lowest overall evaluation loss of 0.3428 and a strong precision of 0.75. Ultimately, this combined strategy matches the Teacher Model’s peak F1-score of 0.81, demonstrating that integrating supervised fine-tuning with knowledge distillation effectively closes the performance gap between the student and teacher models.


\end{document}